%% file: ConformalSurfel.tex
\documentclass[10pt,runningheads]{llncs}
\usepackage[latin1]{inputenc}  % accents 8 bits dans le source

\usepackage{amsmath,amssymb}

\newcommand{\href}[1]{\texttt{#1}}
\usepackage[dvipdfm]{graphicx,color} % without Whizzy

\begin{document}
 \author{Christian \textsc{Mercat}} 
 \authorrunning{Ch. \textsc{Mercat}}                                
\institute{I3M, Université Montpellier 2 c.c. 51\\ 
F-34095 Montpellier cedex 5 France\\ \email{mercat@math.univ-montp2.fr}} 

\titlerunning{Discrete Complex 3D-Surfaces}%
\title{Discrete Complex Structure on Surfel Surfaces}
\maketitle

\begin{abstract}
  This paper defines a theory of conformal parametrization of digital
  surfaces made of surfels equipped with a normal vector. The main
  idea is to locally project each surfel to the tangent plane,
  therefore deforming its aspect-ratio. It is a generalization of the
  theory known for polyhedral surfaces. The main difference is that
  the conformal ratios that appear are no longer real in general. It
  yields a generalization of the standard Laplacian on weighted
  graphs.
\end{abstract}
\section{Introduction}
Conformal parametrization of surfaces is a useful technique in image
processing. The key notion is to identify the tangent plane of a
surface to the field of the complex numbers in a consistent way. It
allows to give a meaning to the notion of angles of two crossing paths
on the surface, or equivalently to the notion of \emph{small circles}
around a point. A surface with such a complex structure is called a
\emph{Riemann} surface. A conformal, holomorphic or analytic function
between two Riemann surfaces is a function that preserves angles and
small circles.

\begin{figure}[h]
  \centering
\scalebox{1.2}{\input{dessins/polygons}}
\caption[ ]{The discrete version of the map $z\mapsto z^3$ in the
  hexagonal/triangular case.}
  \label{fig:z3}
\end{figure}
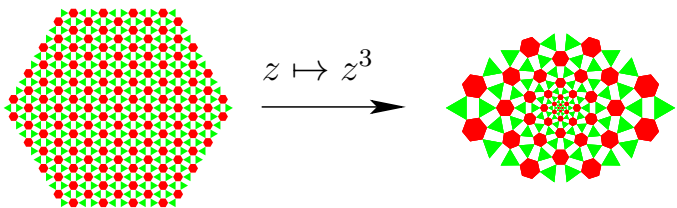

This notion has been tremendously successful in mathematics and
engineering; an aim of this paper is to define its discrete
counterpart in the context of the surfel surfaces, defining for
example discrete polynomials (see Fig.~\ref{fig:z3}).  It is a crucial
notion in \emph{texture mapping} for example: consider a particular
animal skin that is known to contain small patterns of a given
shape, like round disks; if this texture is rendered in a way that
stretches these patterns into ovals, the picture will be wrongly
interpreted by the viewer, the \emph{distortion} being understood as
conveying an information of \emph{tilt} of the underlying surface.

The technique has many other uses, like vector fields on surfaces,
surface \emph{remeshing}, surface recognition or surface
interpolation.  One of its main features is its \emph{rigidity}: the
Riemann mapping theorem tells you that a surface topologically
equivalent to a disc can be conformally mapped to the unit disc,
\emph{uniquely} up to the choice of three points. In this way,
surfaces which are very different can be mapped to the same space
where their features can be compared. There is much less freedom than
in the case of harmonic mapping for example (kernel of the Laplacian,
see \eqref{eq:Laplacian}), which depends on many
arbitrary choices, which are too numerous or too sensitive in many
cases. This technique is surprisingly robust to changes in the bulk of
the surface and the dependency on the boundary conditions can be
relaxed as well~\cite{DMA02}, putting rigidity where the data is
meaningful.

This technique has been widely used in the polyhedral surfaces
community~\cite{DMA02,DKT06,KSS06,GY02,GY03,GYSGP03,JWYG04}. \textbf{In
this paper} we will describe its adaptation to the case of digital
surfaces made of surfels, square boundaries of voxels in
$\mathbf{Z}^3$ that constitute a simple combinatorial surface where
each edgel belongs to one or two surfels. We develop in this article
the theory and algorithms needed for an actual implementation on
computers of these notions to the context of surfel surfaces.

The additional information that we use in order to give the digital
surface a conformal structure is the data of the normal direction~\cite{LMR96,Len97,Mal02}.

In the first section~\ref{sec:conf}, we will present how this
information encodes a non real discrete conformal structure. Then, in
Sec.~\ref{sec:deRham} we will recall elements of de Rham cohomology
and apply them in Sec.~\ref{sec:Hodge} to define a discrete Hodge star
from this conformal structure. How this leads to a theory of discrete
Riemann surfaces will be explained in
Sec.~\ref{sec:Real}-\ref{sec:NonReal}, first recalling the real case,
then generalizing to the complex case, which is the main technical
result of this paper.

\section{Previous work}

Conformal maps are present in a lot of areas. They are called
analytic, holomorphic, meromorphic or monodriffic. They are related to
harmonic maps because they are in particular complex harmonic. In the
context of discrete geometry processing, they were used mainly by the
polyhedral community, for example Desbrun and al.
in~\cite{DMA02,DKT06} or Gu and Yau~\cite{GY02,GY03,GYSGP03,JWYG04}.
Circle packings have been used as well to approximate conformal
maps, see~\cite{KSS06} and references therein.

\section{Conformal structure}\label{sec:conf}

We show here how a surfel surface equipped with normals defines a
discrete conformal structure and gives a geometric interpretation to
holomorphic maps.

A discrete object is a set of points in $\mathbf{Z}^3$, each center of
its Voronoi cell called \emph{voxel}.  A voxel is a cube of unit side,
its six faces are called\emph{ surfels}.  A digital surface $\Sigma$
made of surfels is a connected set of surfels. We will restrict
ourselves to surfaces such that every edge in $\Sigma$ belongs to at
most two surfels~\cite{BM03}. The edges that belong to only one surfel
are called
\emph{boundary edges}. Let us call (the indices stand for dimensions)
$(\lozenge_0,\lozenge_1,\lozenge_2)$ the sets of vertices, edges and
surfels of this cellular decomposition $\lozenge$ of the surface
$\Sigma$.

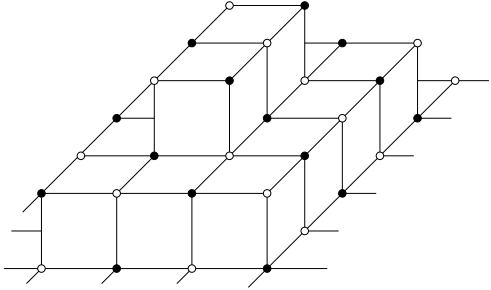
\begin{figure}[h]
\centering
  \scalebox{0.5}{\input{dessins/pyram}}
\caption{A surfel surface}\label{fig:pyram}
\end{figure}

Note that this cellular decomposition is \emph{bipartite}: there
exists a bicoloring of its vertices, that can be colored whether black
or white, no adjacent vertices have the same color. We consider the
surfels diagonals, their end points share the same color, forming \emph{dual}
black and white diagonals.

We call $\Gamma$ the $1$-skeleton graph, whose vertices $\Gamma_0$ are
given by the black vertices and its edges $\Gamma_1$ by the black
diagonals. It can be completed into a cellular decomposition of the
surface by adding faces $\Gamma_2$ for each white vertex. Similarly we
define its \emph{Poincaré dual} $\Gamma^*$ composed of the white
vertices, white diagonals and faces associated with black vertices.We
will refer to $\Gamma$ as the \emph{primal} (black) graph and to
$\Gamma^*$ as its \emph{dual} (white) graph. The $k$-cells of $\Gamma$
are nothing else than the $2-k$ cells of $\Gamma^*$:
$\Gamma_k\eqsim\Gamma^*_{2-k}$.

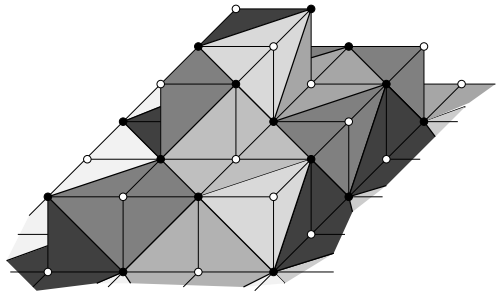
\begin{figure}[h]
\centering
  \scalebox{0.5}{\input{dessins/pyramDual}}
  \caption[ ]{The cellular decomposition $\Gamma$ associated with black vertices}
  \label{fig:pyramDual}
\end{figure}

The data of a \emph{normal direction} at each surfel is a broadly used
feature of digital surfaces~\cite{LMR96,Len97,Mal02}. This normal
might come from a digital scanner, or be computed from the digital
surface itself by various means on which we won't elaborate. These
consistent normals give an orientation to the surface.

This normal is used to project a given surfel comprising the four
vertices $(x,y,x',y')$ to the \emph{local tangent plane}.  This
projection deforms the square into a parallelogram. Its diagonals are
sent to segments which are no longer orthogonal in general. We
identify the tangent plane with the complex plane, up to the choice of
a similitude. We call $Z$ this local map from the cellular
decomposition to the complex numbers. Each diagonal $(x,x')$ and
$(y,y')$ is now seen as a complex number $Z(x')-Z(x)$, resp.
$Z(y')-Z(y)$.

\begin{figure}[h]
\centering
  \scalebox{0.5}{\input{dessins/pyramProj}}
  \caption[ ]{A surfel projected onto the local tangent plane}
  \label{fig:pyramProj}
\end{figure}
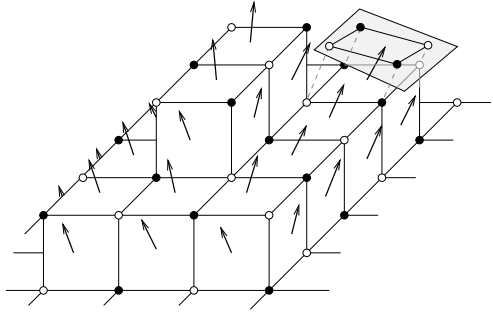

For example we can project the standard digital plane of cubes
associated with $P_0: x+y+z=0$ onto this (constant tangent) plane
$P_0$ and get the following rhombi pattern.
\begin{figure}[h]
  \centering
  \scalebox{0.25}{\input{dessins/surfelHexaTri}}
  \caption[ ]{The digital plane $x+y+z=0$ projected. Note that $\Gamma$
    is the hexagonal lattice, $\Gamma^*$ its triangular dual.}
  \label{fig:TriHex}
\end{figure}
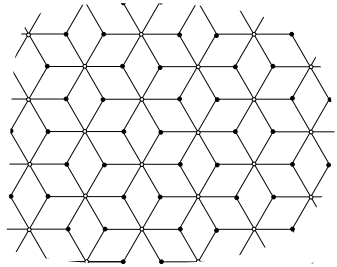

We then associate to each diagonal $(x,x')\in\Gamma_1$ the (possibly
infinite) complex ratio $i\,\rho$ of the dual diagonal by the primal
diagonal, as complex numbers.
$$i\,\rho(x,x'):=\frac{Z(y')-Z(y)}{Z(x')-Z(x)}$$
 This ratio clearly does not depend on
the choice of identification between the tangent plane and the field
of complex numbers.

We will prefer the number $\rho$ to the ratio $i\, \rho$ and still
call it abusively the ratio. This number does not depend on the
orientation of the edge, for $\rho(x,x')=\rho(x',x)$. The number
$\rho$ is real whenever the normal is orthogonal to (at least) one of
the diagonals, that is to say, when the two diagonals are orthogonal
when projected to the tangent plane. We will call this eventuality
\emph{the real case}. It is so, for example, in the standard plane case in
Fig.~\ref{fig:TriHex} where its value is the constant
$\rho_{\text{hex}}=\tan(\frac\pi 6)=1/\sqrt{3}$. The flat square grid
$\mathbb{Z}^2$ is associated with the constant $1$. Large or small
values away from $1$ appear whenever the surfels are flattened away
from the square aspect ratio. The complex valuess appear when the
surfel is slanted away from the orthogonal conformation.
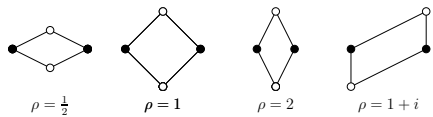
\begin{figure}[h]
  \centering
  \scalebox{0.5}{\input{dessins/rhos}}
  \caption[ ]{Several surfel conformations and the associated ratio $\rho$.}
  \label{fig:rhos}
\end{figure}

We call this data of a graph $\Gamma$, whose edges are equipped with a
complex number $\rho$, a \emph{discrete conformal structure}, or a
\emph{discrete Riemann surface}.

We equip the dual edge $(y,y')\in\Gamma^*_1$ of the complex constant
$\rho(e^*)=1/\rho(e)$. In the example Fig.~\ref{fig:TriHex}, its value
is the constant $\rho_{\text{tri}}=\tan(\frac\pi
3)=\sqrt{3}=1/\rho_{\text{hex}}$.

We define a map $f:\lozenge_0\to\mathbb{C}$ as \emph{discrete
holomorphic} with respect to $\rho$ if and only if, it respects the
ratio for each surfel: $$\forall\, (x,y,x',y')\in\lozenge_2,\;
{f(y')-f(y)}=i\,\rho(x,x')\left({f(x')-f(x)}\right).$$

In the continuous complex analytic theory, a holomorphic function $f$
is a complex function from the complex plane to itself, which is
complex differentiable; that is to say, it is recognized by the fact
that its action on a neighborhood around a point $z_0\in\mathbf{C}$ is
locally a similitude $z\mapsto a\,z + b$ where $b=f(z_0)$ and
$a=f'(z_0)$, the derivative of $f$ at $z_0$ with respect to the
complex variable $z$. It sends little circles to little circles.

In the same fashion, a discrete conformal map sends each
quadrilateral surfel to another quadrilateral whose diagonals have
the same ratio $\rho$ and it can be pictured as sending polygons that form
the double $\Lambda$ into similar polygons that still fit
together. For example, the hexagonal and triangular lattices, hidden
in Fig.~\ref{fig:TriHex}, can actually be drawn on the same
picture by simply joining the middles of the edgels together.
The hexagons and triangles are simply both shrunk by a factor half and
fit together at their vertices. A discrete conformal map is
recognizable in the sense that it sends each of these regular hexagons
and equilateral triangles to polygons of the same shape, touching with
the same combinatorics. The discrete derivative is encoded in how much
each polygon has been inflated or shrunk, and rotated.

\section{de Rham Cohomology}\label{sec:deRham}

In the continuous theory of surfaces, the notion of complex structure
relies on the existence of an operator on $1$-forms such that
$*^2=-\text{Id}$, called the \emph{Hodge star}.  In orthonormal local
coordinates $(x,y)$, it is defined on $1$-forms by $*(f\, dx + g\, dy)
= -g\, dx + f\, dy$. The discrete analogous of these local coordinates
are given by these pairs of dual diagonals $(x,x')$ and $(y,y')$.

In order to follow further this analogy, we first have to define the
spaces of discrete functions and discrete forms such as $dx, dy$. This
is done in the theory of de-Rham cohomology~\cite{FK}.

In this section we recall elements of de Rham cohomology, with
functions, boundary operator and forms, in which the notions of
discrete analytic functions take place. The novelty is that we need to
double everything to get the best out of the complex structure.

We define $\Lambda := \Gamma\sqcup\Gamma^*$, disjoint union of the two
dual graphs, that we will call the \emph{double} graph. Its vertices
$\Lambda_0=\lozenge_0$ are the same as the vertices of the surfel
cellular decomposition $\lozenge$, its edges $\Lambda_1$ are the black
and white diagonals, its faces $\Lambda_2$ is a set in bijection with
its vertices. This doubling can look artificial at first sight but is
in fact very useful in practice, allowing for nicer formulae because
$\Lambda$ is self-dual by construction. We could as well define
complex functions as being real on the vertices $\Gamma_0$ and pure
imaginary on the faces $\Gamma_2$, but it is not as practical. The
case is similar to the continuous where functions and $2$-forms are
essentially the same set but treated differently.

The complex of \emph{chains} $C(\Lambda)=C_{0}(\Lambda)\oplus
C_{1}(\Lambda)\oplus C_{2}(\Lambda)$ is the vector space (over
$\mathbb{R}$ the field of reals) spanned by vertices, edges and faces.
It is equipped with a \emph{boundary} operator
$\partial:C_{k}(\Lambda)\to C_{k-1}(\Lambda)$, null on vertices and
fulfilling $\partial^{2}=0$.  The kernel
$\text{ker~}\partial=:Z_{\bullet}(\Lambda)$ of the boundary operator
are the closed chains or \emph{cycles}.  The image of the boundary
operator are the \emph{exact} chains. The \emph{homology} of $\Lambda$
is the space of cycles modulo exact chains. It encapsulates
all there is to know about the topology of the surface.

The dual spaces of forms are called \emph{cochains}, denoted with
upper indices for the dimension, are functions from chains to the
field of complex numbers.
$C^{k}(\Lambda):=\text{Hom}(C_{k}(\Lambda),\mathbb{C})$. The space
$C^0(\Lambda)$ of $0$-forms is the linear span of functions of the vertices,
$1$-forms are functions of the oriented edges, $2$-forms are functions
of the faces. Coupling is denoted by functional and integral notation:
the value of a $1$-form $\alpha$ evaluated on an edge
$(x,x')\in\Lambda_1$ will be denoted by $\int_x^{x'}\alpha$,
similarly, $f(x)$ for a $0$-form on a vertex, $\iint_F \omega$ for a
$2$-form on a face. The dual of the boundary operator is called the
\emph{coboundary} operator $d:C^k(\Lambda)\to C^{k+1}(\Lambda)$,
defined by Stokes formula: $$\int_x^{x'} df:=
f\left(\partial(x,x')\right)=f(x')-f(x),
\qquad \iint\limits_F d\alpha:=\oint\limits_{\partial F}\alpha,
$$ where $\oint$ denotes the circulation of a $1$-form along a closed
contour. A \emph{cocycle} is a closed cochain and we note $\alpha\in
Z^{k}(\Lambda)$. The \emph{cohomology} of $\Lambda$ is the space of
cocycles modulo the exact forms.

We define an exterior wedge product, denoted $\wedge$, for $1$-forms living
either on edges $\lozenge_1$ or on their diagonals $\Lambda_1$, as a
$2$-form living on surfels $\lozenge_2$. The formula for the latter
is:
\begin{align}
  \label{eq:Wedge}
  \iint\limits_{(x,y,x',y')} \alpha\wedge \beta  &:= 
%\int\limits_{(x,y)}\alpha_\lozenge\int\limits_{(y,x')}\beta_\lozenge
%+\int_{(y,x')}\alpha_\lozenge\int_{(x',y')}\beta_\lozenge\\
%&&\qquad
%+\int_{(x',y')}\alpha_\lozenge\int_{(y',x)}\beta_\lozenge
%+\int_{(y',x)}\alpha_\lozenge\int_{(x,y)}\beta_\lozenge\\
%&:= 
\frac12\left( \int\limits_{(x,x')}\alpha\int\limits_{(y,y')}\beta
-\int\limits_{(y,y')}\alpha\int\limits_{(x,x')}\beta\right)
\end{align}
The exterior derivative $d$ is, as it should be, a derivation for the
wedge product, for functions $f, g$ and a $1$-form $\alpha$:
\begin{align*}
  d(f g) &= f\, dg + g\, df,
& d(f\alpha) = df\wedge\alpha+f\, d\alpha.
\end{align*}

\section{Hodge star}\label{sec:Hodge}
After having set the spaces where our operator is going to act, we
show, in this section, how the discrete conformal structure allows to
define a Hodge star, that is to say, an operator verifying
$*^2=-\text{Id}$ on $1$-forms. It breaks forms into holomorphic and
anti-holomorphic parts. Moreover, we show that this decomposition is
robust to flips and that parallel planes have isomorphic
decompositions.

\begin{definition} The \emph{Hodge star} is defined on functions and
  $2$-forms by
\begin{eqnarray}
    *:C^{k}(\Lambda) & \to & C^{2-k}(\Lambda)
    \notag  \\
   \phantom{*:}C^{0}(\Lambda)\ni f & \mapsto & * f : \iint_{F}*f := f(F^{*}),
    \notag  \\
     \phantom{*:}C^{2}(\Lambda)\ni \omega & \mapsto & * \omega : 
     (*\omega)(x) := \iint_{x^{*}}\omega\;,
    \notag  
\end{eqnarray}
and on $1$-forms, on the dual edges $(y,y')=(x,x')^*\in\Lambda_1$,
given the (complex) discrete conformal structure $\rho(x,x')=r
e^{i\theta}$, by
  $$\begin{pmatrix}
\int_x^{x'}{* \alpha}\\ \int_y^{y'}{* \alpha}
\end{pmatrix}
:=
  \frac 1{\cos\theta}
\begin{pmatrix}
-\sin\theta& -\frac 1 r\\ r & \sin\theta
\end{pmatrix}
\begin{pmatrix}
\int_{(x,x')}{\alpha}\\ \int_{(y,y')}{\alpha}
\end{pmatrix}
.
$$
\end{definition}

Notice that the Hodge star is a \emph{real} transformation.  The fact
that it is well defined relies on the fact that two dual diagonals are
associated with inverse numbers: $\rho(y,y')=1/\rho(x,x')=\frac 1 r
e^{-i\theta}$ (using the former notation). It fulfills $*^2=\text{Id}$
for functions and $2$-forms, and $*^2=-\text{Id}$ for $1$-forms:

We have, on the surfel $(x,y,x',y')\in\lozenge_2$, $$ \frac
1{\cos^2\theta}
\begin{pmatrix}
-\sin\theta& -\frac 1 r\\ r & \sin\theta
\end{pmatrix}^2=
  \frac 1{\cos^2\theta}
\begin{pmatrix}
\sin^2\theta-1&0\\ 0 & \sin^2\theta-1
\end{pmatrix}= -I_2.$$

Because of this property, we can define it on the complexified forms
and functions. It breaks the complex $1$-forms into two eigenspaces
associated to eigenvalues $-i$ and $+i$ called respectively type $(1,0)$
and $(0,1)$ forms.

\begin{definition}
  A \emph{holomorphic} form $\alpha\in C^1_{\mathbf{C}}(\Lambda)$ is a
  closed type $(1,0)$ form, that is to say, $\alpha$ is such that $
  d\alpha = 0$ and $* \alpha = -i \alpha$.  An \emph{anti-holomorphic}
  form is a closed type $(0,1)$ form. 

A function $f$ is holomorphic iff its exterior derivative $df$ is as
well holomorphic, that is to say, on a surfel
$(x,y,x',y')\in\lozenge_2$, $f(y')-f(y)=i\;\rho(x,x')\; (f(x')-f(x)).$

  A \emph{meromorphic} form with a \emph{pole} at $F\in\lozenge_2$, is
  a type $(1,0)$ form, closed except on the surfel $F$. Its lack of
  closeness is called its \emph{residue} at $F$.
\end{definition}

Holomorphic and meromorphic functions are tremendously important in
mathematics, they are behind all the keys on a calculator like
polynomials, inversion, cosine, tangent, exponential,
logarithm... This article defines the framework in which their
discrete counterparts take place.

An interesting feature of the theory is its robustness with respect to
local moves. A discrete holomorphic map defined on a discrete Riemann
surface is mapped by a \emph{canonical isomorphism} to the space of
discrete holomorphic maps defined on another discrete Riemann surface
linked to the original one by a series of \emph{flips}. These flips
are called in the context of discrete conformal structures
\emph{electrical moves}. They come in three kinds, the third one being
the flip, the others being irrelevant to our context.

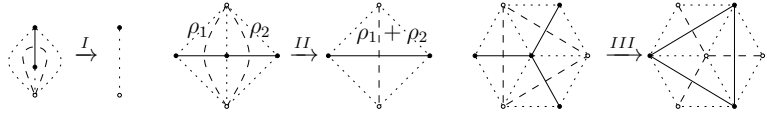
\begin{figure}[htbp]
\begin{center}\scalebox{0.75}{\input{dessins/moves}}
\end{center}
\caption[ ]{The electrical moves.}    \label{fig:moves}
\end{figure}

The third move corresponds to the flip, it is called the \emph{star-triangle
  transformation}.  To three surfels (drawn as dotted lines in
Fig.~\ref{fig:moves}) arranged in a hexagon, whose diagonals form a
triangle of conformal parameters $\rho_{1}$, $\rho_{2}$ and
$\rho_{3}$, one associates a configuration of three other surfels
whose diagonals form a three branched star with conformal parameters
$\rho'_{i}$ (on the opposite side of $\rho_i$) verifying
\begin{equation}
    \rho_{i}\rho'_{i}=\rho_{1}\rho_{2}+\rho_{2}\rho_{3}+\rho_{3}\rho_{1}
    =\frac{\rho'_{1}\,\rho'_{2}\,\rho'_{3}}{\rho'_{1}+\rho'_{2}+\rho'_{3}}.
    \label{eq:rho'rho}
\end{equation}

The value of a holomorphic function at the center of an hexagon is
overdetermined with respect to the six values on the hexagon. These
values have to fulfill a compatibility condition, which are the same
for both hexagons, therefore a holomorphic function defined on a
discrete Riemann surface can be uniquely extended to another surface
differing only by a flip~\cite{BMS05}.

This means in particular that a discrete holomorphic function defined
on the standard plane in~Fig.~\ref{fig:TriHex} can be followed through
all its other parallel deformations and is not sensitive to some added
noise (flips deleting or inserting extra voxels) provided the normal
vector is unchanged with respect to the discrete plane value: the
space of holomorphic functions on these parallel or noisy planes are
in one-to-one correspondence. This theoretical robustness has yet to be
experimentally observed in practice because the normal vectors are not
independent and a noisy plane will have noisy normal vectors as
well. This will be the subject of a forthcoming article.

For real conformal structures, the formulae are simpler, we present
them independently before generalizing them to the complex case.

\section{The Real Case}\label{sec:Real}

We present in this section the formulae in the real case, for the
Hodge star, the Laplacian, the scalar product and the different
energies, Dirichlet and conformal.

When the complex structure is defined by a real ratio $\rho$, for
example in the hexagonal/triangular standard discrete plane case,
then, for each surfel, the Hodge star takes the simpler form
\begin{eqnarray}
    *:C^{k}(\Lambda) & \to & C^{2-k}(\Lambda)
    \notag  \\
   \phantom{*:}C^{0}(\Lambda)\ni f & \mapsto & * f : \iint_{F}*f := f(F^{*}),
    \notag  \\
     \phantom{*:}C^{1}(\Lambda)\ni\alpha & \mapsto & * \alpha :
     \int_{e}*\alpha := -\rho(e^{*})\int_{e^{*}}\alpha,
    \label{eq:*Def}  \\
     \phantom{*:}C^{2}(\Lambda)\ni \omega & \mapsto & * \omega : 
     (*\omega)(x) := \iint_{x^{*}}\omega.
    \notag  
\end{eqnarray}

The endomorphism
$\Delta:=-d*d*-*d*d$ is, in this real case, the usual discrete
\emph{Laplacian}\index{discrete Laplacian}: Its formula on a function
at a vertex $x\in\Gamma_0$ with neighbours $x_1,\ldots,x_V\in\Gamma_0$
is the usual weighted averaged difference:
\begin{equation}\label{eq:Laplacian}
\left(\Delta(f)\right)(x)=
\sum_{k=1}^V\rho(x,x_k)\left(f(x)-f(x_k)\right).
\end{equation}
 The space of \textit{harmonic forms} is defined as its kernel.

 Together with the Hodge star, they give rise, in the compact case, to
 the usual weighted scalar product on $1$-forms:
\begin{equation}
  \label{eq:ScalProd}
  \left(\alpha,\,\beta\right):=\iint_{\lozenge_2} \alpha\wedge
  *\overline\beta = 
(*\alpha,\, *\beta) 
= \overline {\left(\beta,\,\alpha\right)}
=\tfrac12\sum_{e\in\Lambda_1}\rho(e)\int_e\alpha\int_e\overline\beta
\end{equation}

The $\ell^2$ norm of the $1$-form $df$, called the \emph{Dirichlet
  energy} of the function $f$, is the average of the usual Dirichlet
energies on each independant graph
\begin{align}
  E_D(f):=\lVert df\rVert^2 &=
  \left(df,\,df\right)=\frac12\sum_{(x,x')\in\Lambda_1}\rho(x,x')
  \left\lvert f(x') - f(x) \right\rvert^2\label{eq:norm}
  \\
  &=\frac{ E_D(f|_\Gamma)+ E_D(f|_{\Gamma^*})}2 .\notag\end{align} The
conformal energy of a map measures its conformality defect, relating
these two harmonic functions.  A conformal map fulfills the
Cauchy-Riemann equation
\begin{equation}
  \label{eq:CR*}
  *\,df = -i\, df.
\end{equation}
Therefore a quadratic energy whose null functions are the holomorphic
ones is
\begin{equation}
  E_C(f) := \tfrac12\lVert df -i *  df\rVert^2.
\label{eq:EC}
\end{equation}
It is related to the Dirichlet energy through the same formula as in
the continuous case:
\begin{align}
E_C(f) &= \tfrac12\left( df  -i * df,\, df  -i * df\right)
\notag\\
&= \tfrac12\lVert df \rVert^2+\tfrac12\lVert-i * df \rVert^2
+\, \text{Re}(df,\, -i * df)\notag\\
&=
  \lVert df \rVert^2 + \, \text{Im} \iint_{\lozenge_2} df\wedge\overline{df}
\notag\\
&=
E_D(f) - 2 \mathcal{A}(f)
\label{eq:ECEDA}
\end{align}
where the area of the image of the application $f$ in the complex
plane has the same formula 
\begin{equation}
\mathcal{A}(f) = \frac i2
\iint_{\lozenge_2} df\wedge\overline{df}
\label{eq:A}
\end{equation}
as in the continuous case.  For a face $(x,y,x',y')\in\lozenge_2$, the
algebraic area of the oriented quadrilateral
$\Bigl(f(x),f(x'),f(y),f(y')\Bigr)$ is given by
\begin{align*}
  \smash{\iint\limits_{(x,y,x',y')}} df\wedge\overline{df}&=
i\, \text{Im}\left(
(f(x')-f(x))\overline{(f(y')-f(y))}
\right)\\
&=-2 i\mathcal{A}\Bigl(f(x),f(x'),f(y),f(y')\Bigr).
\end{align*}

When a holomorphic reference map $z:\Lambda_0\to\mathbb{C}$ is chosen,
a holomorphic (resp. anti-holomorphic) $1$-form $df$ is, locally on
each pair of dual diagonals, proportional to $dz$, resp. $d\overline z$, so
that the decomposition of the exterior derivative into holomorphic and
anti-holomorphic parts yields $df\wedge\overline{df} = \left(|\partial
  f|^2 +|\overline\partial f|^2\right) dz\wedge d\overline z$ where the
derivatives naturally live on faces.

All these concepts turn into an actual machinery that can be
implemented on computer and a full featured theory mimicking the
theory of Riemann surfaces to a surprisingly far extent~\cite{M01}. In
particular in the rhombic case, the notion of polynomials and
exponentials, differential equations, logarithm (called the Green
function) follow through. A striking difference is the fact that
constants are of dimension two, the constants on the graph $\Gamma$
being independent from the constants on the dual graph $\Gamma^*$.

\section{Non real conformal structure}\label{sec:NonReal}

In this section we give formulae for the scalar product, the Dirichlet
energy and the Laplacian in the complex case. Nothing more is needed in
order to implement computation of global conformal parametrization of
surfel surfaces.

In the complex case, the dual graphs, which are independent in the real case,
are no longer independent and are mixed together.

The first construction that needs an adaptation is the scalar product.
The formula \eqref{eq:ScalProd} still defines a positive definite
scalar product which is preserved by $*$, even though the last
equality must be replaced by a mixed sum over the two dual edges:
\begin{equation}
  \label{eq:ScalProdComp}
  \left(\alpha,\,\beta\right)
=\tfrac12\sum_{e\in\Lambda_1}
\frac{\int_e\alpha}{\text{Re}\left(\rho(e)\right)}
\left(|\rho(e)|^2\int_e\bar\beta+
\text{Im}\left(\rho(e)\right)\int_{e^*}\bar\beta\right).
\end{equation}

The Dirichlet energy mixes the two dual graphs as well:
\begin{equation}
 E_D(f):=\lVert
df\rVert^2
=\frac12\sum_{e\in\Lambda_1}
\frac{\left\lvert f(x') - f(x) \right\rvert^2}{\text{Re}\left(\rho(e)\right)}
\left(|\rho(e)|^2+
\text{Im}\left(\rho(e)\right)\frac{\overline{f(y')-f(y)}}{\overline{f(x')-f(x)}}
\right)
\label{eq:normComp}
\end{equation}

and the Laplacian no longer splits on the two independent dual graphs:
Given $x_0\in\Lambda_0$, with dual face $x_0^*=(y_1,\, y_2,\ldots,\,
y_V)\in\Lambda_2$ and neighbours $x_1,\, x_2,\ldots,\,
x_V\in\Lambda_0$, with dual edges
$(x_0,x_k)^*=(y_k,y_{k+1})\in\Lambda_1$, and $y_{V+1}=y_1$, we have
\begin{equation}
  \label{eq:LaplacianComp}
  \Delta (f) (x_0) = \sum_{k=1}^V \frac{1}{\text{Re}\left(\rho(e)\right)}
\left(|\rho(e)|^2\bigl(f(x_k)-f(x)\bigr)+
\text{Im}\left(\rho(e)\right)\bigl(f(y_{k+1})-f(y_k)\bigr)\right).
\end{equation}

This Laplacian is still real and involves not only the neighbors along
the diagonals of surfels on a weighted graph, which is the star used
for the usual Laplacian, but all the vertices of the surfels that
contains the vertex.

\section{Conclusions and acknowledgments}\label{sec:conclo}

This paper defines the theory of analytic functions on digital
surfaces made of surfels. It required more than a mere adaptation of
the theory known in the polyhedral community since the conformal
parameters are in general not real.

In forthcoming papers, this theory, already put to use in the context
of polyhedral surfaces, will be implemented in the context of surfel
surfaces: The first applications that come to mind are the recognition
of digital surfaces, simple ones and of higher topology through the
computation of their discrete period matrices~\cite{M07}, the analysis
of vector fields on digital surfaces allowed by the Hodge theorem that
decomposes vector fields into rotational and divergence free parts,
the creation of vector fields with given circulation properties, in
general the correct discrete treatment of partial differential
equations on a digital surface which are solved by analytic functions
in the continuous case, like incompressible fluid dynamics for
example.

The author would like to thank the referees for useful comments, the
forthcoming Géométrie Discrète ANR project, in particular Valérie
Berthé (LIRMM) and Rémy Malgouyres (LLAIC) for discussions that
brought the main idea that made this paper possible.

\bibliographystyle{splncs} \bibliography{surface}
\end{document}

%% file: dessins/pyram.tex
\begin{picture}(0,0)%
\includegraphics{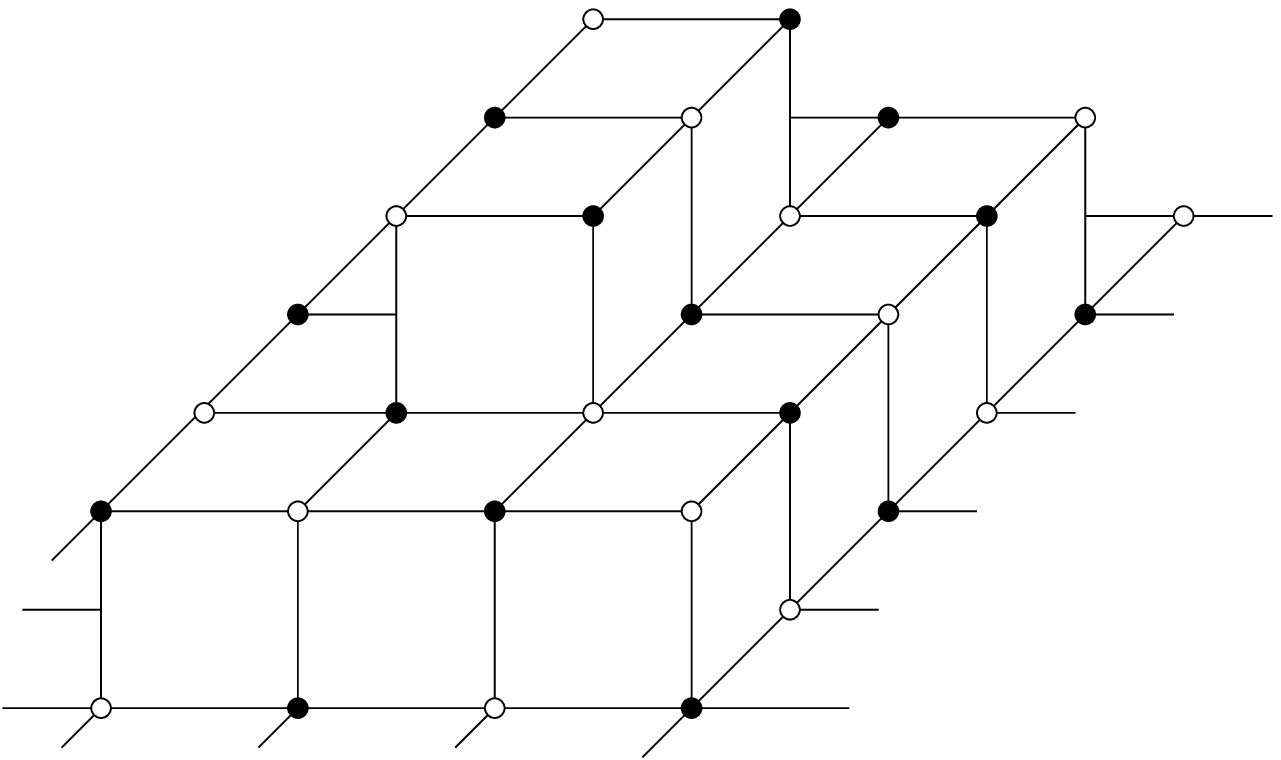}%
\end{picture}%
\setlength{\unitlength}{4144sp}%
\begingroup\makeatletter\ifx\SetFigFont\undefined%
\gdef\SetFigFont#1#2#3#4#5{%
  \reset@font\fontsize{#1}{#2pt}%
  \fontfamily{#3}\fontseries{#4}\fontshape{#5}%
  \selectfont}%
\fi\endgroup%
\begin{picture}(5829,3440)(-461,-2998)
\end{picture}%

%% file: dessins/pyramDual.tex
\begin{picture}(0,0)%
\includegraphics{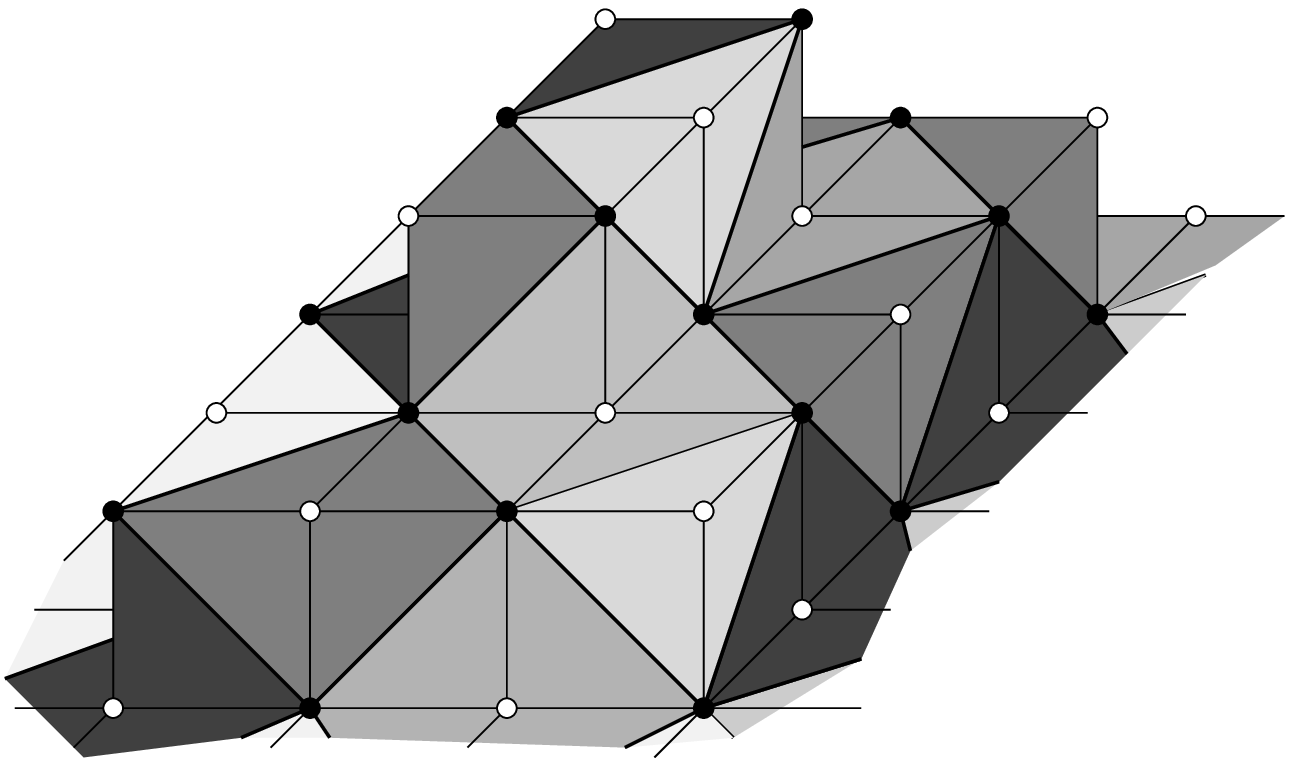}%
\end{picture}%
\setlength{\unitlength}{4144sp}%
\begingroup\makeatletter\ifx\SetFigFont\undefined%
\gdef\SetFigFont#1#2#3#4#5{%
  \reset@font\fontsize{#1}{#2pt}%
  \fontfamily{#3}\fontseries{#4}\fontshape{#5}%
  \selectfont}%
\fi\endgroup%
\begin{picture}(5884,3440)(-516,-2998)
\end{picture}%

%% file: dessins/pyramProj.tex
\begin{picture}(0,0)%
\includegraphics{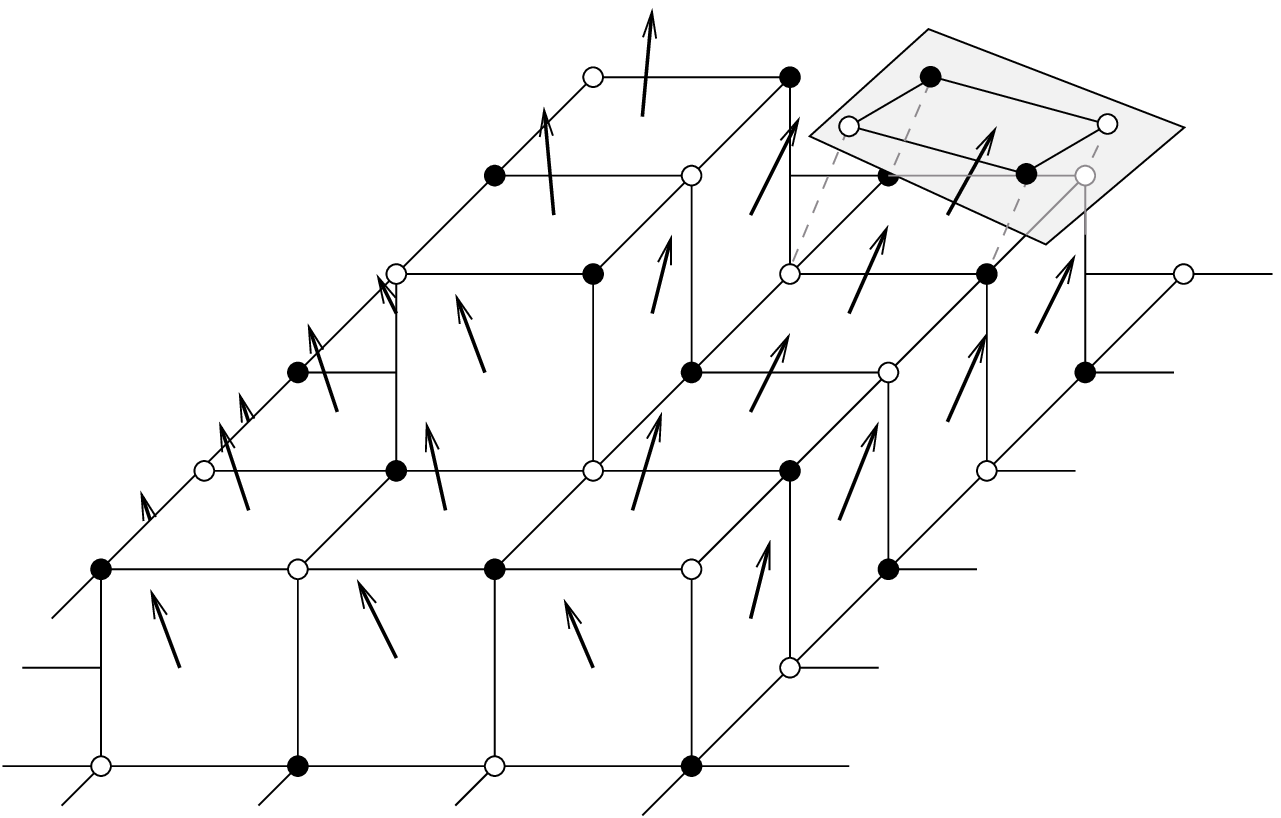}%
\end{picture}%
\setlength{\unitlength}{4144sp}%
\begingroup\makeatletter\ifx\SetFigFont\undefined%
\gdef\SetFigFont#1#2#3#4#5{%
  \reset@font\fontsize{#1}{#2pt}%
  \fontfamily{#3}\fontseries{#4}\fontshape{#5}%
  \selectfont}%
\fi\endgroup%
\begin{picture}(5829,3724)(-461,-2998)
\end{picture}%

%% file: dessins/surfelHexaTri.tex
\begin{picture}(0,0)%
\includegraphics{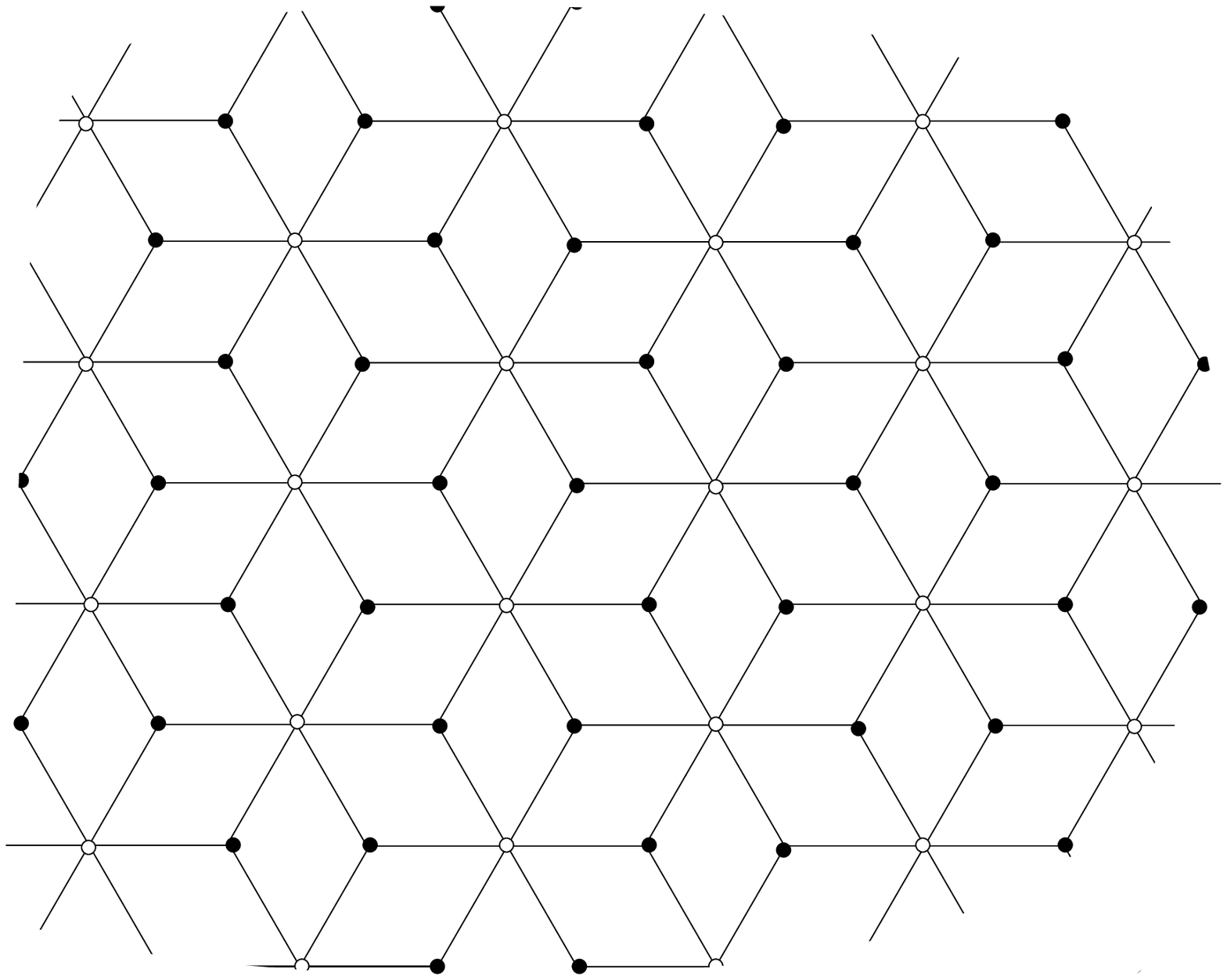}%
\end{picture}%
\setlength{\unitlength}{4144sp}%
\begingroup\makeatletter\ifx\SetFigFont\undefined%
\gdef\SetFigFont#1#2#3#4#5{%
  \reset@font\fontsize{#1}{#2pt}%
  \fontfamily{#3}\fontseries{#4}\fontshape{#5}%
  \selectfont}%
\fi\endgroup%
\begin{picture}(9561,8246)(43,-7568)
\end{picture}%

%% file: dessins/rhos.tex
\begin{picture}(0,0)%
\includegraphics{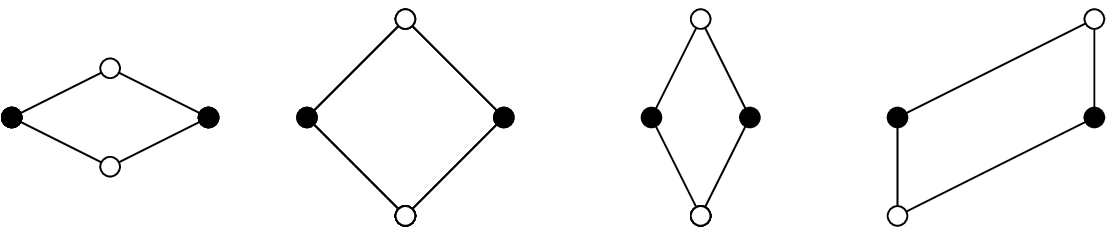}%
\end{picture}%
\setlength{\unitlength}{4144sp}%
\begingroup\makeatletter\ifx\SetFigFont\undefined%
\gdef\SetFigFont#1#2#3#4#5{%
  \reset@font\fontsize{#1}{#2pt}%
  \fontfamily{#3}\fontseries{#4}\fontshape{#5}%
  \selectfont}%
\fi\endgroup%
\begin{picture}(5056,1292)(173,-625)
\put(676,-556){\makebox(0,0)[b]{\smash{{\SetFigFont{12}{14.4}{\rmdefault}{\mddefault}{\updefault}{\color[rgb]{0,0,0}$\rho=\frac12$}%
}}}}
\put(2026,-556){\makebox(0,0)[b]{\smash{{\SetFigFont{12}{14.4}{\rmdefault}{\mddefault}{\updefault}{\color[rgb]{0,0,0}$\rho=1$}%
}}}}
\put(2026,-556){\makebox(0,0)[b]{\smash{{\SetFigFont{12}{14.4}{\rmdefault}{\mddefault}{\updefault}{\color[rgb]{0,0,0}$\rho=1$}%
}}}}
\put(3376,-556){\makebox(0,0)[b]{\smash{{\SetFigFont{12}{14.4}{\rmdefault}{\mddefault}{\updefault}{\color[rgb]{0,0,0}$\rho=2$}%
}}}}
\put(4726,-556){\makebox(0,0)[b]{\smash{{\SetFigFont{12}{14.4}{\rmdefault}{\mddefault}{\updefault}{\color[rgb]{0,0,0}$\rho=1+i$}%
}}}}
\end{picture}%

%% file: dessins/polygons.tex
\begin{picture}(0,0)%
\includegraphics{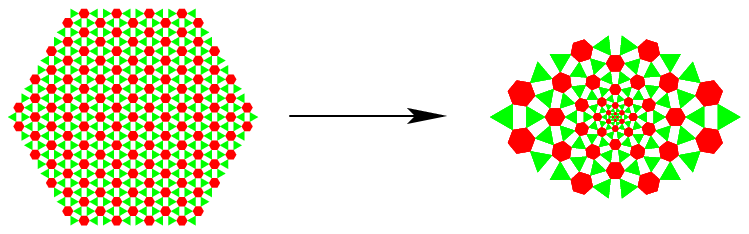}%
\end{picture}%
\setlength{\unitlength}{4144sp}%
\begingroup\makeatletter\ifx\SetFigFont\undefined%
\gdef\SetFigFont#1#2#3#4#5{%
  \reset@font\fontsize{#1}{#2pt}%
  \fontfamily{#3}\fontseries{#4}\fontshape{#5}%
  \selectfont}%
\fi\endgroup%
\begin{picture}(3475,1270)(181,-926)
\put(1531,-151){\makebox(0,0)[lb]{\smash{{\SetFigFont{12}{14.4}{\rmdefault}{\mddefault}{\updefault}{\color[rgb]{0,0,0}$z\mapsto z^3$}%
}}}}
\end{picture}%

%% file: dessins/moves.tex
\begin{picture}(0,0)%
\includegraphics{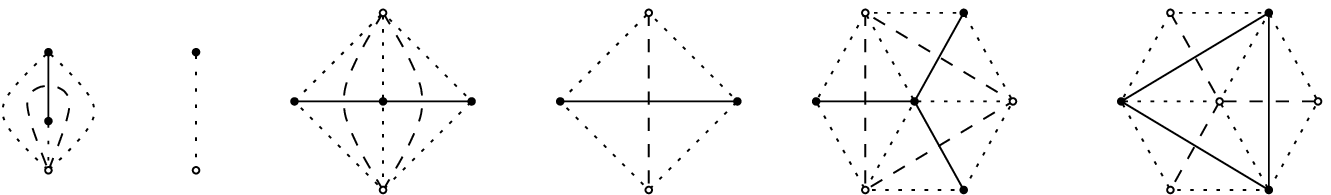}%
\end{picture}%
\setlength{\unitlength}{4144sp}%
\begingroup\makeatletter\ifx\SetFigFont\undefined%
\gdef\SetFigFont#1#2#3#4#5{%
  \reset@font\fontsize{#1}{#2pt}%
  \fontfamily{#3}\fontseries{#4}\fontshape{#5}%
  \selectfont}%
\fi\endgroup%
\begin{picture}(6050,856)(94,-39)
\put(1531,569){\makebox(0,0)[lb]{\smash{{\SetFigFont{12}{14.4}{\rmdefault}{\mddefault}{\updefault}{\color[rgb]{0,0,0}$\rho_1$}%
}}}}
\put(631,344){\makebox(0,0)[lb]{\smash{{\SetFigFont{12}{14.4}{\rmdefault}{\mddefault}{\updefault}{\color[rgb]{0,0,0}$\xrightarrow{I}$}%
}}}}
\put(2026,569){\makebox(0,0)[lb]{\smash{{\SetFigFont{12}{14.4}{\rmdefault}{\mddefault}{\updefault}{\color[rgb]{0,0,0}$\rho_2$}%
}}}}
\put(2341,344){\makebox(0,0)[lb]{\smash{{\SetFigFont{12}{14.4}{\rmdefault}{\mddefault}{\updefault}{\color[rgb]{0,0,0}$\xrightarrow{II}$}%
}}}}
\put(2881,524){\makebox(0,0)[lb]{\smash{{\SetFigFont{12}{14.4}{\rmdefault}{\mddefault}{\updefault}{\color[rgb]{0,0,0}$\rho_1+\rho_2$}%
}}}}
\put(4861,344){\makebox(0,0)[lb]{\smash{{\SetFigFont{12}{14.4}{\rmdefault}{\mddefault}{\updefault}{\color[rgb]{0,0,0}$\xrightarrow{III}$}%
}}}}
\end{picture}%